\documentclass[aps,amssymb,amsmath,
twocolumn,showpacs,superscriptaddress,
groupedaddress,ifmbe]{revtex4-1}

\usepackage{graphicx}
\usepackage{natbib}
\usepackage{dcolumn}
\usepackage{bm}
\usepackage[table,xcdraw]{xcolor}
\usepackage{nth}
\usepackage[ruled,vlined]{algorithm2e}
\usepackage{framed}
\usepackage{listings}
\usepackage[titletoc,title]{appendix}
\usepackage{tikz}
\usepackage{lipsum}
\usepackage{hyperref}
\usepackage{multirow}
\usepackage{nicefrac}
\usepackage[table]{xcolor}
\usepackage{threeparttable}
\usepackage{comment}
\usepackage{dirtytalk}

\hypersetup{
  colorlinks,
  citecolor=red,
  linkcolor=blue,
  urlcolor=blue}

\begin{document}
\title{New Atomic Orbital Functions.\\ Complete and Orthonormal Sets of ETOs with Non$-$integer Quantum Numbers.\\Results for He$-$like atoms. }
\author{Ali Ba{\u g}c{\i}}
\email{abagci@pau.edu.tr}
\affiliation{Computational and Gravitational Physics Laboratory, Department of Physics, Faculty of Science, Pamukkale University, Denizli, Turkey}
\author{Philip E. Hoggan}
\affiliation{Institut Pascal, UMR 6602 CNRS, BP 80026, 63178 Aubiere Cedex, France.}

\begin{abstract}
The Hartree$-$Fock$-$Rothaan equations are solved for He$-$like ions using the iterative self$-$consistent method. New complete and orthonormal sets of exponential$-$type orbitals are employed as the basis. These orbitals satisfy the orthonormality condition for quantum numbers with fractional power. They are solutions of a Schr{\"o}dinger$-$like differential equation derived by the authors. In a recent study conducted for the calculation of the hydrogen atom energy levels, it has been demonstrated that the fractional formalism of the principal and the angular momentum quantum numbers converges to the $1s$ level of the ground state energy of hydrogen atom, obtained from the solution of the standard Schr{\"o}dinger equation. This study examines the effect of fractional values of the quantum numbers for two-electron systems, which is the simplest system with electron correlation effects.
\begin{description}
\item[Keywords]
Exponential$-$type orbitals, Hartree$-$Fock equations, He$-$like atoms
\end{description}
\end{abstract}
\maketitle

\section{Introduction} \label{introduction}
New basis functions for many$-$electron systems expressing solutions to the Dirac equation in the non$-$relativistic limit are presented here. A basis spans Hilbert space and any function in that space can be expressed as a linear combination thereof. One approach to defining basis functions is Sturm$-$Liouville theory. The eigen$-$functions of the Hamiltonian, expressed as a Liouville operator, form a complete set and are suitable as basis functions. The advantages of the present set of eigen$-$functions are enumerated. They are defined from eigen$-$functions of the (relativistic) Dirac equation, in the large $c$ limit.

The non$-$relativistic description of atoms, molecules and solids involves solving the electronic Schr{\"o}dinger equation. Methods for doing this generally separate the electrons, to treat the system as $N$ one$-$electron equations.\\
The Schr{\"o}dinger differential equation for a one$-$electron Hamiltonian has 'hydrogen$-$like' eigen$-$functions \cite{1_Schiff_1968}:
\begin{align}\label{eq:1}
\psi_{nlm}\left(\vec{r}\right)
=R_{n}^{l}\left(r\right)S_{lm}\left(\theta, \varphi\right),
\end{align}
here,
\begin{multline}\label{eq:2}
R_{n}^{l}\left(r\right)
=\sqrt{\left(\frac{2Z}{na_{\mu}}\right)^{3}
\frac{\left(n-l-1\right)!}{2n\left(n+l\right)!}}
e^{-\frac{Zr}{n a_{\mu}}}\left(\frac{2Z}{n a_{\mu}}\right)^{l}
\\
\times L_{n-l-1}^{2l+1}\left( \frac{2Zr}{n a_{\mu}} \right),
\end{multline}
$S_{lm}$ are normalized complex $\left(S_{lm}\equiv Y_{lm}, Y_{lm}^{\ast} \right)$ or real spherical harmonics \cite{2_Stevenson_1965}. $L_{q-p}^{p} \left(x\right)$ are associated Laguerre polynomials \cite{3_Magnus_1966}. $Z$ is the nuclear charge, $a_{\mu}$ is the Bohr radius.\\
The electron probability distribution of the Eq. (\ref{eq:1}) $\vert \psi_{nlm}\left(\vec{r}\right) \vert^{2}$, characterizes the spatial probability measure of electronic states in hydrogen$-$like atoms. This is indeed a measurable electron density. However, hydrogen$-$like eigen$-$functions  do not form a complete basis. The continuum states must expressly be included. This restricts their representation in the corresponding Hilbert space. The issue is addressed by treating the orbital exponent $\zeta$ as a variational parameter not containing n, $\zeta=Z/a_{\mu}$ \cite{4_Hylleraas_1928}.  Such exponents are used in orthonormal eigenfunction basis sets including Lambda functions \cite{5_Shull_1959} and Coulomb$-$Sturmians \cite{6_Rotenberg_1962} These are eigenfunctions of a Schr{\"o}dinger-like equation where the Coulomb potential is scaled such that the exponent does not contain quantum number $n$. 

The present non$-$relativistic complete orthonormal basis (Ba{\u g}c{\i}$-$Hoggan ETOs, BH$-$ETOs) \cite{7_Bagci_2023} comprises exponential$-$type orbitals, which arise as solutions to a Schr{\"o}dinger$-$like differential equation obtained as the mathematically rigorous non$-$relativistic limit of the Dirac equation. The relativistic treatment of the hydrogen atom \cite{8_Grant_2007} as formulated through the Dirac equation \cite{9_Dirac_1928} thus, provides a foundation for understanding the emergence of Ba{\u g}c{\i}$-$Hoggan ETOs. This is achieved by adapting their governing equation to the relativistic Dirac$-$like counterpart  \cite{10_Bagci_2025} (and references therein).

The first objective of this work is to analyze the applicability of BH$-$ETOs to treat multi$-$electron atoms and molecules, with an initial focus on fundamental systems that lack analytical solutions, such as two$-$electron helium$-$like ions, using the standard self$-$consistent field (SCF) procedure for the matrix formulation of the Hartree$-$Fock equations, namely the Hartree$-$Fock$-$Roothaan (HFR) equations \cite{11_Roothaan_1951}. From a quantum chemical standpoint, accounting for electron correlation evidently requires further examination. As elements of basic Hilbert space, the adoption of non$-$integer principal quantum number basis functions in non$-$relativistic electronic structure calculations based on the independent$-$particle model contributes to convergence, at the cost of increasing the number of parameters to be optimized, yet does not reflect any physically meaningful structure. Ba{\u g}c{\i}$-$Hoggan ETOs provide an extension of the Coulomb$-$Sturmian or Shull Lambda functions to non$-$integer principal quantum numbers, thereby defining functions that lie outside the subspace spanned by the integer $n$ family, which forms an orthonormal basis in the basic Hilbert space.\\
The second objective is the numerical validation of the aforementioned application. Such numerical validation will hold significant relevance for the discussions regarding basis set construction methods through the functions with non$-$integer principal quantum numbers. This will be presented in the next paper of this series.

\section{On the Origin of Ba{\u g}c{\i}$-$Hoggan ETOs}\label{OrgBH}
The following relationship for the operator $\left(\vec{\sigma}.\hat{\vec{p}} \right)$ contributes to the solution of the Dirac equation in a spherically symmetric Coulomb potential,
\begin{align}\label{eq:3}
\left(\vec{\sigma}.\hat{\vec{p}} \right)=
-i\vec{\sigma}.\hat{r}\left[ i\hat{r}.\hat{\vec{p}}-
\frac{\vec{\sigma}.\left( \hat{r} \times \hat{\vec{p}} \right)}{r} \right],
\end{align}
here, $\vec{\sigma}=\left(\sigma_{x}, \sigma_{y}, \sigma_{z} \right)$ are the Pauli spin matrices and $\hat{\vec{p}}$ is the momentum operator. The Eq. (\ref{eq:3}) leads to \cite{12_Szmytkowski_2007},
\begin{multline}\label{eq:4}
\left(\vec{\sigma}.\hat{\vec{p}} \right)f\left(r\right)\Omega_{\kappa \mu}\left(\theta, \varphi\right)=
\\
i\left[
\frac{d f\left(r\right)}{dr}+\frac{\kappa+1}{r}f\left(r\right)
\right]\Omega_{\kappa \mu}\left(\theta, \varphi\right),
\end{multline}
with, $f\left(r\right)$ an arbitrary radial function and $\Omega_{\kappa \mu}\left(\theta, \varphi\right)$ are spin one$-$half spherical spinor harmonics \cite{13_Varshalovich_1988}. They are given in matrix form as \cite{12_Szmytkowski_2007},
\begin{align}\label{eq:5}
\Omega_{\kappa\mu}\left( \theta,\varphi \right)
\begin{pmatrix}
sgn\left( -\kappa \right)
\sqrt{\frac{\kappa+1/2-\mu}{2\kappa + 1}}
Y_{l\mu-1/2}\left(\theta, \varphi\right)
\\
\sqrt{\frac{\kappa+1/2+\mu}{2\kappa + 1}}
Y_{l\mu+1/2}\left(\theta, \varphi\right).
\end{pmatrix}
\end{align}

The upper$-$ and lower$-$components of the radial parts of the Dirac equation eigen$-$functions for hydrogen$-$like atoms are given respectively by \cite{14_Johnson_2004},
\begin{multline}\label{eq:6}
f_{n\kappa}^{L}\left(r\right)=
\mathcal{N}_{n\kappa}\sqrt{1+W_{n\kappa}}\left(2\zeta r\right)^{\gamma}e^{-\zeta r}
\\
\times
\left\lbrace
\left(N_{n\kappa}-\kappa\right)
F\left[-\left(n-\kappa\right), 2\gamma+1, 2\zeta r\right]
\right.
\\
\left.
-\left(n-\kappa\right)F\left[-\left(n-\kappa\right)+1, 2\gamma+1, 2\zeta r \right]
\right\rbrace ,
\end{multline}
\begin{multline}\label{eq:7}
f_{n\kappa}^{S}\left(r\right)=
\mathcal{N}_{n\kappa}\sqrt{1-W_{n\kappa}}\left(2\zeta r\right)^{\gamma}e^{-\zeta r}
\\
\times
\left\lbrace
\left(N_{n\kappa}-\kappa\right)
F\left[-\left(n-\kappa\right), 2\gamma+1; 2\zeta r\right]
\right.
\\
\left.
+\left(n-\kappa\right)
F\left[-\left(n-\kappa\right)+1, 2\gamma+1; 2\zeta r \right]
\right\rbrace ,
\end{multline}
where $\mathcal{N}_{n\kappa}$ are normalization constants, $n_{r}=n-k=n-\vert \kappa \vert$, $\kappa=\pm 1, \pm 2, \pm 3, ...$,
\begin{align*}
 \zeta=Z/\sqrt{\left(\alpha Z\right)^{2}+\left[\left(n-k\right)+\gamma \right]^{2}}.   
\end{align*}
$W_{n\kappa}$ are used to characterize the discrete bound$-$state solution. The $F\left[a,b;z\right]$ are confluent hyper$-$geometric functions of the first kind. They are related to the generalized Laguerre polynomials \cite{3_Magnus_1966} as follows:
\begin{align}\label{eq:8}
L_{q-p}^{p}\left(x\right)=\frac{\left(p+1\right)_{q-p}}{\Gamma\left(q-p+1\right)}
F\left[-\left(q-p\right), p+1;x\right],
\end{align}
$\left(x\right)_{n}$ is the Pochhammer symbol. $L-$spinors \cite{15_Grant_2000} are thus, derived from relativistic analogues of Coulomb$-$Sturmians with fractional vakues of:
\begin{align*}
\gamma=\sqrt{\kappa^{2}-\frac{Z^2}{c^2}},
\end{align*}
$c$ speed of light and,
\begin{align*}
 N_{n\kappa}=\sqrt{n_{r}^{2}+2n_{r}\gamma+\kappa^{2}}.   
\end{align*}
In the non$-$relativistic limit $\left(c\rightarrow \infty \right)$ the lower component of the Dirac equation solution goes to zero, while the upper component converges to the Schr{\"o}dinger equation eigen$-$functions. The non$-$relativistic limit is derived from the following properties of the generalized Laguerre polynomials \cite{3_Magnus_1966},
\begin{align}\label{eq:9}
L_{q-p}^{p}\left(x\right)=L_{q-p}^{p+1}\left(x\right) -L_{q-p-1}^{p+1}\left(x\right),
\end{align}
\begin{multline}\label{eq:10}
xL_{q-p}^{p+1}\left(x\right)
\\
=\left(p+1\right)L_{q-p}^{p}\left(x\right)
-\left(q-p+1\right)L_{q-p+1}^{p}\left(x\right)
\end{multline}

A recent investigation by authors \cite{7_Bagci_2023} has established the need for an intermediate form, termed \textit{transitional Laguerre} polynomials, between generalized Laguerre polynomials and standard Laguerre polynomials. Although the authors work \cite{7_Bagci_2023} offers a thorough analysis on the subject, this necessity becomes explicit when Eqs. (\ref{eq:6}) and (\ref{eq:7}) are expressed in their Rodrigues forms. Ba{\u g}c{\i}$-$Hoggan complete and orthonormal sets of exponential$-$type orbitals, together with their corresponding differential equation, have been derived to eliminate this mathematical requisite for the Schr{\"o}dinger equation for one-electron atoms. The resulting Schr{\"o}dinger$-$type differential equation has been then extended into a Dirac$-$type equation that accounts for relativistic effects \cite{10_Bagci_2025}. In the weighted Hilbert space $L_{r^{\alpha}}\left(\mathbb{R}^{3} \right)$, (square integrable) BH$-$ETOs are written:
\begin{multline}\label{eq:11}
R_{n^{\ast}l^{\ast}}^{\alpha\nu}\left( \zeta, r \right)
\\
=\mathcal{N}_{n^{\ast}l^{\ast}}^{\alpha\nu}\left(\zeta\right)
\left(2\zeta r\right)^{l^{\ast}+\nu-1}e^{-\zeta r}
L_{n^{\ast}-l^{\ast}-\nu}^{2l^{\ast}+2\nu-\alpha}\left(2\zeta r\right),
\end{multline}
$\left\lbrace n^{\ast},l^{\ast} \right\rbrace \in \mathbb{R}$ and $0 < \nu \leq 1$.

\begin{table*}[t!]
\renewcommand{\arraystretch}{1.15}
\caption{\label{tab:table1}
Comparison between Hartree$-$Fock limit values from Ref. \cite{24_King_2018} and those obtained using BH$-$ETOs in HFR equations with integer, fractional principal quantum numbers for ground state energy of He atom. The single$-$zeta approximation is used.
}
\begin{threeparttable}
\begin{ruledtabular}
\begin{tabular}{cccc}
$q$ & $\left\lbrace n^{*},\zeta \right\rbrace$ & $E$ & $\Delta E$ 
\\ \hline
1
&
\begin{tabular}[c]{@{}l@{}}
\begin{tabular}{cc}
0.95505 74100 & 1.61172 47267 
\end{tabular}
\\
\begin{tabular}{cc}
1.00000 00000 & 1.68750 00336 
\end{tabular}
\\
{}
\end{tabular}
&
\begin{tabular}[c]{@{}l@{}}
\textbf{-2.8}\underline{54}20 84970 26459 
\\
\textbf{-2.8}\underline{47}65 62499 99999
\\
\textbf{-2.8}\underline{47}65 6250 \tnote{a}
\end{tabular}
&
\begin{tabular}[c]{@{}l@{}}
-0.00747 14985 85780
\\
-0.01402 37456 12240
\\
{}
\end{tabular}
\\ \cline{1-2}
3
&
\begin{tabular}[c]{@{}l@{}}
\begin{tabular}{cc}
0.99640 49719 & 1.90086 41592 
\end{tabular}
\\
\begin{tabular}{cc}
1.00000 00000 & 1.92088 17379 
\end{tabular}
\\
{}
\end{tabular}
&
\begin{tabular}[c]{@{}l@{}}
\textbf{-2.8616}\underline{0 8}9244 82740
\\
\textbf{-2.86}\underline{15}9 00546 63452
\\
\textbf{-2.86}\underline{15}9 00547 \tnote{a}
\end{tabular}
&
\begin{tabular}[c]{@{}l@{}}
-0.00007 10711 29499
\\
-0.00008 99409 48787
\\
{}
\end{tabular}
\\ \cline{1-2}
5
&
\begin{tabular}[c]{@{}l@{}}
\begin{tabular}{cc}
1.00029 08686 & 1.97530 10175
\end{tabular}
\\
\begin{tabular}{cc}
1.00000 00000 & 1.96654 42246
\end{tabular}
\\
{}
\end{tabular}
&
\begin{tabular}[c]{@{}l@{}}
\textbf{-2.86167 9}\underline{75}52 21257
\\
\textbf{-2.86167 9}\underline{67}52 54109
\\
\textbf{-2.86167 9}\underline{67}53 \tnote{a}
\end{tabular}
&
\begin{tabular}[c]{@{}l@{}}
-0.00000 02403 90982
\\
-0.00000 03203 58130
\\
{}
\end{tabular}
\\ \cline{1-2}
7
&
\begin{tabular}[c]{@{}l@{}}
\begin{tabular}{cc}
1.00036 69189 & 1.80745 34011
\end{tabular}
\\
\begin{tabular}{cc}
1.00000 00000 & 1.75451 96860
\end{tabular}
\\
{}
\end{tabular}
&
\begin{tabular}[c]{@{}l@{}}
\textbf{-2.86167 9}\underline{87}39 27975
\\
\textbf{-2.86167 9}\underline{76}34 43967
\\
\textbf{-2.86167 9}\underline{76}34 \tnote{a}
\end{tabular}
&
\begin{tabular}[c]{@{}l@{}}
-0.00000 01216 84264
\\
-0.00000 02321 68272
\\
{}
\end{tabular}
\\ \cline{1-2}
9
&
\begin{tabular}[c]{@{}l@{}}
\begin{tabular}{cc}
1.00012 22784 & 2.34173 53791
\end{tabular}
\\
\begin{tabular}{cc}
1.00000 00000 & 2.34173 53791
\end{tabular}
\\
{}
\end{tabular}
&
\begin{tabular}[c]{@{}l@{}}
\textbf{-2.86167 99}\underline{71}0 02725
\\
\textbf{-2.86167 99}\underline{63}1 94484
\\
\textbf{-2.86167 99}\underline{63}2 \tnote{a}
\end{tabular}
&
\begin{tabular}[c]{@{}l@{}}
-0.00000 00246 09514
\\
-0.00000 00324 17755
\\
{}
\end{tabular}
\\ \cline{1-2}
11
&
\begin{tabular}[c]{@{}l@{}}
\begin{tabular}{cc}
1.00002 81353 & 2.48399 76643
\end{tabular}
\\
\begin{tabular}{cc}
1.00000 00000 & 2.49861 95567
\end{tabular}
\\
{}
\end{tabular}
&
\begin{tabular}[c]{@{}l@{}}
\textbf{-2.86167 999}\underline{42} 33439
\\
\textbf{-2.86167 999}\underline{39} 22505
\\
\textbf{-2.86167 999}\underline{39} \tnote{a}
\end{tabular}
&
\begin{tabular}[c]{@{}l@{}}
-0.00000 00013 78799
\\
-0.00000 00016 89734
\\
{}
\end{tabular}
\\ \cline{1-2}
13
&
\begin{tabular}[c]{@{}l@{}}
\begin{tabular}{cc}
1.00000 57480 & 2.60899 59701
\end{tabular}
\\
\begin{tabular}{cc}
1.00000 00000 & 2.60899 40744
\end{tabular}
\end{tabular}
&
\begin{tabular}[c]{@{}l@{}}
\textbf{-2.86167 9995}\underline{5 4}7598
\\
\textbf{-2.86167 9995}\underline{5 3}6816
\end{tabular}
&
\begin{tabular}[c]{@{}l@{}}
-0.00000 00000 64641
\\
-0.00000 00000 75423
\end{tabular}
\\ \cline{1-2}
15
&
\begin{tabular}[c]{@{}l@{}}
\begin{tabular}{cc}
1.00000 10837 & 2.69199 57648
\end{tabular}
\\
\begin{tabular}{cc}
1.00000 00000 & 2.69199 51252
\end{tabular}
\end{tabular}
&
\begin{tabular}[c]{@{}l@{}}
\textbf{-2.86167 99956} \underline{09}557
\\
\textbf{-2.86167 99956} \underline{09}217
\end{tabular}
&
\begin{tabular}[c]{@{}l@{}}
-0.00000 00000 02682
\\
-0.00000 00000 03022
\end{tabular}
\\ \cline{1-2}
17
&
\begin{tabular}[c]{@{}l@{}}
\begin{tabular}{cc}
1.00000 01648 & 2.75321 27675
\end{tabular}
\\
\begin{tabular}{cc}
1.00000 00000 & 2.75321 15754
\end{tabular}
\end{tabular}
&
\begin{tabular}[c]{@{}l@{}}
\textbf{-2.86167 99956 12}\underline{14}4
\\
\textbf{-2.86167 99956 12}\underline{13}6
\end{tabular}
&
\begin{tabular}[c]{@{}l@{}}
-0.00000 00000 00095
\\
-0.00000 00000 00103
\end{tabular}
\\ \cline{1-2}
19
&
\begin{tabular}[c]{@{}l@{}}
\begin{tabular}{cc}
1.00000 00152 & 2.79056 99778
\end{tabular}
\\
\begin{tabular}{cc}
1.00000 00000 & 2.79056 97936
\end{tabular}
\end{tabular}
&
\begin{tabular}[c]{@{}l@{}}
\textbf{-2.86167 99956 1223}6
\\
\textbf{-2.86167 99956 1223}6
\end{tabular}
&
\begin{tabular}[c]{@{}l@{}}
-0.00000 00000 00003
\\
-0.00000 00000 00003
\end{tabular}
\\ \cline{1-2}
21
&
\begin{tabular}[c]{@{}l@{}}
\begin{tabular}{cc}
1.00000 00000 & 2.78668 07895
\end{tabular}
\\
{}
\end{tabular}
&
\begin{tabular}[c]{@{}l@{}}
\hspace{19mm}
\textbf{-2.86167 99956 12238 8}2624
\\
\hspace{19mm}
\textbf{-2.86167 99956 12238 8}7877 55437 \tnote{b}
\end{tabular}
&
\begin{tabular}[c]{@{}l@{}}
{}
\\
{}
\end{tabular}

\end{tabular}
\begin{tablenotes}
\item[a] Ref. \cite{21_Guseinov_2008}
\item[b] Ref. \cite{24_King_2018}
\item {$\Delta E = E_{\text{\cite{24_King_2018}}} - E$; upper value: $n^{\ast} \in \mathbb{R}^{+}$ (non$-$integer), lower value: $n^{\ast} \in \mathbb{N}^{+}$ (integer).}
\end{tablenotes}
\end{ruledtabular}
\end{threeparttable}
\end{table*}

\begin{table*}[t!]
\renewcommand{\arraystretch}{1.15}
\caption{\label{tab:table2}
Comparison between Hartree$-$Fock limit values from Ref. \cite{25_Hatano_2020} and those obtained using BH$-$ETOs in HFR equations with integer, fractional principal quantum numbers for some ground state energy of He$-$like atoms. The single$-$zeta approximation is used.
}
\begin{threeparttable}
\begin{ruledtabular}
\begin{tabular}{ccccc}
Atom & $q$ & $\left\lbrace n^{*},\zeta \right\rbrace$ & $E$ & $\Delta E$
\\ \cline{2-5}
\multirow{10}{*}{$C^{4+}$}
&
1
&
\begin{tabular}[c]{@{}l@{}}
\begin{tabular}{cc}
0.98586 97270 & 5.60714 09549
\end{tabular}
\\
\begin{tabular}{cc}
1.00000 00000 & 5.68749 43415 
\end{tabular}
\\
{}
\end{tabular}
&
\begin{tabular}[c]{@{}l@{}}
\textbf{-32.3}\underline{54}37 12869 83980 
\\
\textbf{-32.3}\underline{47}65 62499 67981
\\
\textbf{-32.3}\underline{47}65 6250 \tnote{a}
\end{tabular}
&
\begin{tabular}[c]{@{}l@{}}
-0.00682 15887 34034
\\
-0.01353 66257 50028
\\
{}
\end{tabular}
\\ \cline{2-3}
&
5
&
\begin{tabular}[c]{@{}l@{}}
\begin{tabular}{cc}
1.00023 26328 & 6.58259 14486
\end{tabular}
\\
\begin{tabular}{cc}
1.00000 00000 & 6.48600 47166
\end{tabular}
\end{tabular}
&
\begin{tabular}[c]{@{}l@{}}
\textbf{-32.36119 2}\underline{28}22 80852
\\
\textbf{-32.36119} \underline{17}529 98584
\end{tabular}
&
\begin{tabular}[c]{@{}l@{}}
-0.00000 05934 37158
\\
-0.00000 11227 19425
\end{tabular}
\\ \cline{2-3}
&
9
&
\begin{tabular}[c]{@{}l@{}}
\begin{tabular}{cc}
1.00002 17048 & 8.12824 99572
\end{tabular}
\\
\begin{tabular}{cc}
1.00000 00000 & 8.12824 62732
\end{tabular}
\end{tabular}
&
\begin{tabular}[c]{@{}l@{}}
\textbf{-32.36119 28}\underline{65}6 39943
\\
\textbf{-32.36119 28}\underline{63}2 22802
\end{tabular}
&
\begin{tabular}[c]{@{}l@{}}
-0.00000 00100 78066
\\
-0.00000 00124 95207
\end{tabular}
\\ \cline{2-3}
&
13
&
\begin{tabular}[c]{@{}l@{}}
\begin{tabular}{cc}
1.00000 04595 & 8.86048 06636
\end{tabular}
\\
\begin{tabular}{cc}
1.00000 00000 & 8.86047 84737
\end{tabular}
\end{tabular}
&
\begin{tabular}[c]{@{}l@{}}
\textbf{-32.36119 28757 1}\underline{29}84
\\
\textbf{-32.36119 28757 1}\underline{23}10
\end{tabular}
&
\begin{tabular}[c]{@{}l@{}}
-0.00000 00000 05022
\\
-0.00000 00000 05702
\end{tabular}
\\ \cline{2-3}
&
17
&
\begin{tabular}[c]{@{}l@{}}
\begin{tabular}{cc}
0.99999 99876 & 9.22708 08285
\end{tabular}
\\
\begin{tabular}{cc}
1.00000 00000 & 9.22707 69959
\end{tabular}
\\
{}
\end{tabular}
&
\begin{tabular}[c]{@{}l@{}}
\textbf{-32.36119 28757 1801}\underline{1}
\\
\textbf{-32.36119 28757 1801}\underline{1}
\\
\textbf{-32.36119 28757 1801} \tnote{b}
\end{tabular}
&
\begin{tabular}[c]{@{}l@{}}
-0.00000 00000 00000
\\
-0.00000 00000 00000
\\
{}
\end{tabular}
\\ \hline
\multirow{10}{*}{$Ne^{8+}$}
&
1
&
\begin{tabular}[c]{@{}l@{}}
\begin{tabular}{cc}
0.99577 03703 & 9.68749 33691 
\end{tabular}
\\
\begin{tabular}{cc}
1.00000 00000 & 9.68749 33691 
\end{tabular}
\\
{}
\end{tabular}
&
\begin{tabular}[c]{@{}l@{}}
\textbf{-93.8}\underline{51}05 27744 71363 
\\
\textbf{-93.8}\underline{47}65 62499 56032
\\
\textbf{-93.8}\underline{47}65 6250  \tnote{a}
\end{tabular}
&
\begin{tabular}[c]{@{}l@{}}
-0.01006 07447 59964
\\
-0.01345 72692 75300
\\
{}
\end{tabular}
\\ \cline{2-3}
&
5
&
\begin{tabular}[c]{@{}l@{}}
\begin{tabular}{cc}
1.00015 30908 & 11.12420 77877
\end{tabular}
\\
\begin{tabular}{cc}
1.00000 00000 & 10.94499 54026
\end{tabular}
\end{tabular}
&
\begin{tabular}[c]{@{}l@{}}
\textbf{-93.86111} \underline{27}481 67037
\\
\textbf{-93.86111} \underline{20}900 73099
\end{tabular}
&
\begin{tabular}[c]{@{}l@{}}
-0.00000 07710 64299
\\
-0.00000 14291 58238
\end{tabular}
\\ \cline{2-3}
&
9
&
\begin{tabular}[c]{@{}l@{}}
\begin{tabular}{cc}
1.00001 22850 & 13.73897 12652
\end{tabular}
\\
\begin{tabular}{cc}
1.00000 00000 & 13.73896 81773
\end{tabular}
\end{tabular}
&
\begin{tabular}[c]{@{}l@{}}
\textbf{-93.86111 35}\underline{09}6 95662
\\
\textbf{-93.86111 35}\underline{07}5 07432
\end{tabular}
&
\begin{tabular}[c]{@{}l@{}}
-0.00000 00095 35678
\\
-0.00000 00117 23905
\end{tabular}
\\ \cline{2-3}
&
13
&
\begin{tabular}[c]{@{}l@{}}
\begin{tabular}{cc}
1.00000 01648 & 14.92464 36051
\end{tabular}
\\
\begin{tabular}{cc}
1.00000 00000 & 14.92465 07874
\end{tabular}
\end{tabular}
&
\begin{tabular}[c]{@{}l@{}}
\textbf{-93.86111 35192} \underline{28}248
\\
\textbf{-93.86111 35192} \underline{27}865
\end{tabular}
&
\begin{tabular}[c]{@{}l@{}}
-0.00000 00000 03067
\\
-0.00000 00000 03463
\end{tabular}
\\ \cline{2-3}
&
17
&
\begin{tabular}[c]{@{}l@{}}
\begin{tabular}{cc}
0.99999 99877 & 15.58872 38283
\end{tabular}
\\
\begin{tabular}{cc}
1.00000 00000 & 15.52017 32012
\end{tabular}
\end{tabular}
&
\begin{tabular}[c]{@{}l@{}}
\textbf{-93.86111 35192 313}\underline{37}
\\
\textbf{-93.86111 35192 313}\underline{38}
\end{tabular}
&
\begin{tabular}[c]{@{}l@{}}
-0.00000 00000 00000
\\
-0.00000 00000 00000
\end{tabular}
\\ \cline{2-3}
&
19
&
\begin{tabular}[c]{@{}l@{}}
\begin{tabular}{cc}
1.00000 00000 & 16.32134 13694
\end{tabular}
\\
{}
\end{tabular}
&
\begin{tabular}[c]{@{}l@{}}
\hspace{3mm}
\textbf{-93.86111 35192 313}\underline{38}
\\
\hspace{3mm}
\textbf{-93.86111 35192 3134}(4) \tnote{b}
\end{tabular}
&
\begin{tabular}[c]{@{}l@{}}
{}
\\
{}
\end{tabular}
\\ \hline
\multirow{3.5}{*}{$Mg^{10+}$}
&
17
&
\begin{tabular}[c]{@{}l@{}}
\begin{tabular}{cc}
1.00000 00000 & 18.64231 39958
\end{tabular}
\end{tabular}
&
\begin{tabular}[c]{@{}l@{}}
\textbf{-136.61109 44329 22}\underline{59}6
\end{tabular}
&
\begin{tabular}[c]{@{}l@{}}
{}
\end{tabular}
\\ \cline{2-3}
&
19
&
\begin{tabular}[c]{@{}l@{}}
\begin{tabular}{cc}
1.00000 00000 & 18.84274 33245
\end{tabular}
\\
{}
\end{tabular}
&
\begin{tabular}[c]{@{}l@{}}
\textbf{-136.61109 44329 22}\underline{59}6
\\
\textbf{-136.61109 44329 2260} \tnote{b}
\end{tabular}
&
\begin{tabular}[c]{@{}l@{}}
{}
\\
{}
\end{tabular}
\\ \hline
\multirow{3.5}{*}{$Si^{12+}$}
&
17
&
\begin{tabular}[c]{@{}l@{}}
\begin{tabular}{cc}
1.00000 00000 & 21.75330 84629
\end{tabular}
\end{tabular}
&
\begin{tabular}[c]{@{}l@{}}
\textbf{-187.36108 09759 309}\underline{37}
\end{tabular}
&
\begin{tabular}[c]{@{}l@{}}
{}
\end{tabular}
\\ \cline{2-3}
&
19
&
\begin{tabular}[c]{@{}l@{}}
\begin{tabular}{cc}
1.00000 00000 & 21.75330 84629
\end{tabular}
\\
{}
\end{tabular}
&
\begin{tabular}[c]{@{}l@{}}
\hspace{3mm}
\textbf{-187.36108 09759 309}\underline{37}
\\
\hspace{3mm}
\textbf{-187.36108 09759 3094}(6) \tnote{b}
\end{tabular}
&
\begin{tabular}[c]{@{}l@{}}
{}
\\
{}
\end{tabular}
\\ \hline
\multirow{3.5}{*}{$S^{14+}$}
&
17
&
\begin{tabular}[c]{@{}l@{}}
\begin{tabular}{cc}
1.00000 00000 & 24.85507 73444
\end{tabular}
\end{tabular}
&
\begin{tabular}[c]{@{}l@{}}
\textbf{-246.11107 09777 93}\underline{25}1
\end{tabular}
&
\begin{tabular}[c]{@{}l@{}}
{}
\end{tabular}
\\ \cline{2-3}
&
19
&
\begin{tabular}[c]{@{}l@{}}
\begin{tabular}{cc}
1.00000 00000 & 24.85507 73444
\end{tabular}
\\
{}
\end{tabular}
&
\begin{tabular}[c]{@{}l@{}}
\textbf{-246.11107 09777 93}\underline{25}1
\\
\textbf{-246.11107 09777 9326} \tnote{b}
\end{tabular}
&
\begin{tabular}[c]{@{}l@{}}
{}
\\
{}
\end{tabular}
\\ \hline
\multirow{3.5}{*}{$Ar^{16+}$}
&
17
&
\begin{tabular}[c]{@{}l@{}}
\begin{tabular}{cc}
1.00000 00000 & 27.94897 82761
\end{tabular}
\end{tabular}
&
\begin{tabular}[c]{@{}l@{}}
\textbf{-312.86106 32568 006}\underline{21}
\end{tabular}
&
\begin{tabular}[c]{@{}l@{}}
{}
\end{tabular}
\\ \cline{2-3}
&
19
&
\begin{tabular}[c]{@{}l@{}}
\begin{tabular}{cc}
1.00000 00000 & 27.94897 82761
\end{tabular}
\\
{}
\end{tabular}
&
\begin{tabular}[c]{@{}l@{}}
\textbf{-312.86106 32568 006}\underline{21}
\\
\textbf{-312.86106 32568 0063} \tnote{b}
\end{tabular}
&
\begin{tabular}[c]{@{}l@{}}
{}
\\
{}
\end{tabular}

\end{tabular}
\begin{tablenotes}
\item[a] Ref. \cite{22_Guseinov_2008}
\item[b] Ref. \cite{25_Hatano_2020}
\item {$\Delta E = E_{\text{\cite{25_Hatano_2020}}} - E$; upper value: $n^{\ast} \in \mathbb{R}^{+}$ (non$-$integer), lower value: $n^{\ast} \in \mathbb{N}^{+}$ (integer).}
\end{tablenotes}
\end{ruledtabular}
\end{threeparttable}
\end{table*}

\section{Progress in the Method of Computation} \label{AMComput}
The matrix representation for two$-$electron atomic systems is solved using combined HFR theory \cite{16_Guseinov_2007}. In this theory the energy expectation value is given by,
\begin{align}\label{eq:12}
E\left(LS\right)
=2\sum_{i}^{n}f_{i}h_{i}
+\sum_{ijkl}^{n}\left( 2A^{ij}_{kl}J^{ij}_{kl}
-B^{ij}_{kl}K^{ij}_{kl}
\right).
\end{align}
$E\left(LS\right)$ given in the Eq. (\ref{eq:12}) is based on the assumption that the energy is the average expectation value for all degenerate total orthonormal sets of multi$-$determinental wave functions. The coupling projection coefficients for closed shell systems are accordingly determined by,
\begin{align}\label{eq:13}
A^{ij}_{kl}=B^{ij}_{kl}=f_{i}f_{k}\delta_{ij}\delta_{kl}
\end{align}
$f_{i}$ is the fractional occupancy of shell $i$ and the elements of matrices $h_{i},J^{ij}_{kl},K^{ij}_{kl}$ are obtained through computation of the one$-$electron, two$-$electron Coulomb and exchange integrals, respectively. Below, the linear combination of atomic orbitals method (LCAO) is used to solve the matrix form of HFR equations, the two$-$electron integrals arising in energy expectation value are given as,
\begin{multline}\label{eq:14}
J^{\alpha pq}_{rs}=\int\int
\Bigl(
\psi_{p}^{\alpha \nu_{1}^{\ast}}\left(x_{1}\right)
\psi_{r}^{\alpha \nu_{2}^{\ast}}\left(x_{2}\right)
\frac{1}{r_{21}}
\Bigr.
\\
\Bigl.
\times \psi_{q}^{\alpha \nu_{1}}\left(x_{1}\right)
\psi_{s}^{\alpha \nu_{2}}\left(x_{2}\right)
\Bigr)
dV_{1}dV_{2},
\end{multline}
\begin{multline}\label{eq:15}
K^{\alpha pq}_{rs}=\int\int
\Bigl(
\psi_{p}^{\alpha \nu_{1}^{\ast}}\left(x_{1}\right)
\psi_{r}^{\alpha \nu_{2}^{\ast}}\left(x_{2}\right)
\frac{1}{r_{21}}
\Bigr.
\\
\Bigl.
\times \psi_{s}^{\alpha \nu_{1}}\left(x_{1}\right)
\psi_{q}^{\alpha \nu_{2}}\left(x_{2}\right)
\Bigr)
dV_{1}dV_{2}.
\end{multline}
Here, indices $\alpha, \nu$ indicate that the BH$-$ETOs are used in LCAO. In conclusion, the generalized eigenvalue equation to be solved is given by,
\begin{align}\label{eq:16}
\sum_{q}\left(
\hat{F}^{\nu i}_{pq}-\epsilon^{\nu}_{i} S_{pq}^{\nu}
\right)C_{qi}^{\nu}=0.
\end{align}
The BH$-$ETOs are used as atomic orbitals in the LCAO method. Hence, the criteria below should be satisfied,
\begin{multline}\label{eq:17}
\lim_{q\rightarrow \infty} 
\left[
\sum_{q}\left(
\hat{F}^{\nu i}_{pq}-\epsilon^{\nu}_{i} S_{pq}^{\nu}
\right)C_{qi}^{\nu}
\right]
\\
= \sum_{q}\left(
\hat{F}^{i}_{pq}-\epsilon_{i} S_{pq}
\right)C_{qi} ,
\end{multline}
$\left(q\rightarrow \infty \Rightarrow \nu \rightarrow 1 \right)$. This paper is structured to numerically verify Eq. (\ref{eq:17}). The HFR equations, following the standard formalism presented in Equation (\ref{eq:12}), are utilized. Conversely, the BH$-$ETOs offer a solution to the generalized Kepler problem in quantum mechanics.  This necessitates an improvement of the HFR formalism derived from the Schr{\"o}dinger equation. It is a nontrivial research topic that lies beyond the scope of the present paper.

Dropping the restrictions on quantum numbers is known to trigger a higher computational cost in solution of Eq. (\ref{eq:16}). This follows from the fact that two$-$electron integrals are expressed in terms of higher transcendental functions. These functions lack closed$-$form representations because their associated differential equations have power series solutions with expansions that are non$-$analytic at the origin. In another recent study \cite{17_Bagci_2024}, for the evaluation of two$-$electron integrals involving higher transcendental functions the first author introduced a bi$-$directional method, complemented by hyper$-$radial functions. The hyper$-$radial functions facilitate a reformulation of integrals containing these transcendental functions, effectively eliminating the need for their explicit computation or reliance on infinite power series expansions. The use of basis sets with fractional quantum numbers in quantum chemical calculations was first proposed by Parr and Joy \cite{18_Parr_1957}. They hypothetically suggested dropping the restriction on quantum numbers (specifically, the principal quantum number) of Slater$-$type orbitals. The historical development and theoretical framework of this subject have been comprehensively discussed by the authors in \cite{19_Bagci_2020}. Subsequently, by refining the work of Infeld and Hull \cite{20_Infeld_1951}, it was demonstrated that the differential equation governing the motion of an electron around a nucleus naturally incorporates fractional quantum numbers. Moreover, the Slater$-$type orbitals with a fractional$-$order principal quantum number (NSTOs) are obtained by considering the highest power of $r$ in the solution of such a differential equation (BH$-$ETOs). The transformations between BH$-$ETOs and NSTOs are given by \cite{7_Bagci_2023},
\begin{align}\label{eq:18}
\psi_{n^{\ast}l^{\ast}m^{\ast}}^{\alpha \nu}\left(\zeta, \vec{r}\right)
\sum_{n^{\prime \ast}=l^{\ast}+\nu}^{n^{\ast}}
a_{n^{\ast}n^{\prime \ast}}^{\alpha \nu l^{\ast}}
\chi_{n^{\prime \ast}l^{\ast}m^{\ast}}\left(\zeta, \vec{r} \right),
\end{align}
\begin{align}\label{eq:19}
\chi_{n^{\ast}l^{\ast}m^{\ast}}\left( \zeta, \vec{r} \right)
\sum_{n^{\prime \ast}=l^{\ast}+\nu}^{n^{\ast}}
\bar{a}_{n^{\ast}n^{\prime \ast}}^{\alpha \nu l^{\ast}}
\psi_{n^{\prime \ast}l^{\ast}m^{\ast}}^{\alpha \nu}\left( \zeta, \vec{r} \right).
\end{align}
The coefficients for the transformations between BH$-$ETOs, NSTOs are given as:
\begin{multline}\label{eq:120}
a^{\alpha\nu l^{*}}_{n^{*}{n'}^{*}}
=\left(-1\right)^{{n'}^{*}-l^{*}-\nu}
\\
\times \Bigg[
\dfrac{\Gamma \left({n'}^{*}+l^{*}+\nu+1\right)}{\left(2n^{*}\right)^{\alpha}\Gamma\left( {n'}^{*}+l^{*}+\nu+1-\alpha\right)} \Bigg. \\
\times \Bigg.
F_{{n'}^{*}+l^{*}+\epsilon-\alpha}\left(n^{*}+l^{*}+\nu-\alpha\right)
\Bigg.
\\ F_{{n'}^{*}-l^{*}-\nu}\left(n^{*}-l^{*}-\nu\right)
F_{{n'}{*}-l^{*}-\nu}\left(2{n'}^{*}\right)
\Bigg]^{1/2},
\end{multline}
\begin{multline}\label{eq:21}
\tilde{a}^{\alpha\nu l^{*}}_{n^{*}{n'}^{*}}
=\left(-1\right)^{{n'}^{*}-l^{*}-\nu}
\\
\times \Bigg[\dfrac{\left(2{n'}^{*}\right)^\alpha \Gamma\left( n^{*}+l^{*}+\nu+1-\alpha \right)}{\Gamma \left( n^{*}+l^{*}+\nu+1\right)F_{n^{*}-l^{*}-\nu}\left(2n^{*}\right)}
\Bigg.\\
\times F_{{n'}^{*}+l^{*}+\nu-\alpha}\left(n^{*}+l^{*}+\nu-\alpha\right) \Bigg.
\\
\times F_{{n'}^{*}-l^{*}-\nu}\left(n^{*}-l^{*}-\nu\right)
\Bigg]^{1/2}.
\end{multline}

This summarizes our examination of electron repulsion integral evaluation methodologies. The Eqs. (\ref{eq:18}, \ref{eq:19}) demonstrate that obtaining solutions for these integrals over the NSTOs would constitute a sufficient approach. The Laplace expansion of Coulomb interactions remains applicable in this context. Consequently, the radial component of the two$-$electron integrals may be formulated analytically in terms of hyper$-$geometric functions as \cite{17_Bagci_2024},
\begin{multline}\label{eq:22}
R_{n^{\ast}n^{\prime \ast}}\left( \zeta, \zeta^{\prime} \right)
=\frac{\Gamma\left( n^{\ast}+n^{\prime \ast}+1 \right)}{\left( \zeta+\zeta^{\prime} \right)^{n^{\ast}+n^{\prime \ast}+1}}
\\
\times
\left\lbrace
\frac{1}{n^{\ast}+L+1}
\right.
\\
\left.
\times {}_{2}F\bigg[1, n^{\ast}+n^{\prime \ast}+1, n^{\ast}+L+2; \frac{\zeta}{\zeta+\zeta^{\prime}} \bigg]
\right.
\\
\left.
+\frac{1}{n^{\prime \ast}+L+1}
\right.
\\
\left.
\times {}_{2}F\bigg[1, n^{\ast}+n^{\prime \ast}+1, n^{\prime \ast}+L+2; \frac{\zeta^{\prime}}{\zeta+\zeta^{\prime}} \bigg]
\right\rbrace
\end{multline}
Using the recurrence relations obtained via the bi$-$directional method, complemented by hyper$-$radial functions, these integrals can be evaluated with a high degree of accuracy and significantly reduced computational cost. A more in$-$depth analysis of progress in the computational method falls beyond the scope of this paper. Those interested in further details could consult \cite{17_Bagci_2024}.
\begin{table*}[t]
\renewcommand{\arraystretch}{1.15}
\caption{\label{tab:table3}%
Values of ground state energies $\left(E\right)$ and virial ratios for some iso$-$electronic series of $He$ atom in double$-$zeta approximation
}
\begin{threeparttable}
\begin{ruledtabular}
\begin{tabular}{ccccc}
Atom & $n^{\ast}_{1s},n^{\ast}_{1s^{\prime}}$ & $\zeta_{1s},\zeta_{1s^{\prime}}$  & $E_{BH-ETOs}$ & Virial (ratio)
\\ \hline
$He$
& 
\begin{tabular}[c]{@{}l@{}}
0.98207
\\
1.01316
\end{tabular}
&
\begin{tabular}[c]{@{}l@{}}
2.85100
\\
1.45434
\end{tabular}
&
\begin{tabular}[c]{@{}l@{}}
\textbf{-2.86167} \underline{3561}3 56041 \tnote{a}
\\
\textbf{-2.86167} \underline{3561} \tnote{b}
\\
\textbf{-2.86167 9996} \tnote{c}
\\Confirmed \tnote{d}
\end{tabular}
&
\begin{tabular}[c]{@{}l@{}}
2.00000 03357 64544
\\
{}
\\
{}
\\
{}
\end{tabular}
\\ \cline{4-5}
$Li^{+}$ 
& 
\begin{tabular}[c]{@{}l@{}}
0.98393
\\
1.00735
\end{tabular}
&
\begin{tabular}[c]{@{}l@{}}
4.51797
\\
2.45011
\end{tabular}
&
\begin{tabular}[c]{@{}l@{}}
\textbf{-7.23641} \underline{26517} 05340 \tnote{a}
\\
\textbf{-7.23641} \underline{2652} \tnote{b}
\\
\textbf{-7.23641 5201} \tnote{c}
\\
Confirmed \tnote{d}
\end{tabular} &
\begin{tabular}[c]{@{}l@{}}
1.99999 91396 44960
\\
{}
\\
{}
\\
{}
\end{tabular}
\\ \cline{4-5}
$Be^{2+}$
& 
\begin{tabular}[c]{@{}l@{}}
1.00854
\\
0.99806
\end{tabular}
&
\begin{tabular}[c]{@{}l@{}}
6.31803
\\
3.44652
\end{tabular}
&
\begin{tabular}[c]{@{}l@{}}
\textbf{-13.61129} \underline{76325} 31102 \tnote{a}
\\
\textbf{-13.61129} \underline{7633} \tnote{b}
\\
\textbf{-13.61129 943} \tnote{c}
\\
Confirmed \tnote{d}
\end{tabular} &
\begin{tabular}[c]{@{}l@{}}
1.99999 99542 43029
\\
{}
\\
{}
\\
{}
\end{tabular}
\\ \cline{4-5}
$B^{3+}$
& 
\begin{tabular}[c]{@{}l@{}}
1.01686
\\
0.99703
\end{tabular}
&
\begin{tabular}[c]{@{}l@{}}
4.44558
\\
9.67733
\end{tabular}
&
\begin{tabular}[c]{@{}l@{}}
\textbf{-21.98623} \underline{30137} 78689 \tnote{a}
\\
\textbf{-21.98623} \underline{3014} \tnote{b}
\\
\textbf{-21.98623 447} \tnote{c}
\\
Confirmed \tnote{d}
\end{tabular} &
\begin{tabular}[c]{@{}l@{}}
1.99999 96687 96194
\\
{}
\\
{}
\\
{}
\end{tabular}
\\ \cline{4-5}
$C^{4+}$
& 
\begin{tabular}[c]{@{}l@{}}
1.0115956
\\
0.9983954
\end{tabular}
&
\begin{tabular}[c]{@{}l@{}}
9.7372126
\\
5.4446241
\end{tabular} 
&
\begin{tabular}[c]{@{}l@{}}
\textbf{-32.36119} \underline{15178} 36514 \tnote{a}
\\
\textbf{-32.36119} \underline{1518} \tnote{b}
\\
\textbf{-32.36119 288} \tnote{c}
\\ \textbf{-32.36119} 16252 43004  \tnote{d}
\end{tabular} &
\begin{tabular}[c]{@{}l@{}}
2.00000 02731 42579
\\
{}
\\
{}
\\
1.99999 99772 247326
\end{tabular}
\\ \cline{4-5}
$N^{5+}$
& 
\begin{tabular}[c]{@{}l@{}}
0.97801
\\
1.00405
\end{tabular}
&
\begin{tabular}[c]{@{}l@{}}
11.08902
\\
6.44519
\end{tabular} 
&
\begin{tabular}[c]{@{}l@{}}
\textbf{-44.73616} \underline{19756} 80457 \tnote{a}
\\
\textbf{-44.73616} \underline{1976} \tnote{b}
\\
\textbf{-44.73616 396} \tnote{c}
\\
Confirmed \tnote{d}
\end{tabular} &
\begin{tabular}[c]{@{}l@{}}
1.99999 97148 55625
\\
{}
\\
{}
\\
{}
\end{tabular}
\\ \cline{4-5}
$O^{6+}$
& 
\begin{tabular}[c]{@{}l@{}}
1.0103130
\\
0.9989763
\end{tabular}
&
\begin{tabular}[c]{@{}l@{}}
13.1032833
\\
7.4433857
\end{tabular}
&
\begin{tabular}[c]{@{}l@{}}
\textbf{-59.11114} \underline{0946}2 44556 \tnote{a}
\\
\textbf{-59.11114} \underline{0946} \tnote{b}
\\
\textbf{-59.11114 270} \tnote{c}
\\
\textbf{-59.11114} 15076 13016 \tnote{d}
\end{tabular} &
\begin{tabular}[c]{@{}l@{}}
2.00000 02268 42462
\\
{}
\\
{}
\\
1.99998 79097 43548
\end{tabular}
\\ \cline{4-5}
$F^{7+}$
& 
\begin{tabular}[c]{@{}l@{}}
1.0091610
\\
0.9992397
\end{tabular}
&
\begin{tabular}[c]{@{}l@{}}
14.7939255
\\
8.4431392
\end{tabular}
&
\begin{tabular}[c]{@{}l@{}}
\textbf{-75.48612} \underline{5169}1 61307 \tnote{a}
\\
\textbf{-75.48612} \underline{5169} \tnote{b}
\\
\textbf{-75.48612 641} \tnote{c}
\\
\textbf{-75.48612} 52046 70313 \tnote{d}
\end{tabular} &
\begin{tabular}[c]{@{}l@{}}
2.00000 01952 47180
\\
{}
\\
{}
\\
1.99999 99480 32315
\end{tabular}
\\ \cline{4-5}
$Ne^{8+}$
& 
\begin{tabular}[c]{@{}l@{}}
1.0103352
\\
0.9992194
\end{tabular}
&
\begin{tabular}[c]{@{}l@{}}
16.5078302
\\
9.4428999
\end{tabular}
&
\begin{tabular}[c]{@{}l@{}}
\textbf{-93.86111} \underline{18305} 86235 \tnote{a}
\\
\textbf{-93.86111} \underline{1831} \tnote{b}
\\
\textbf{-93.86111 352} \tnote{c}
\\
\textbf{-93.86111} 23550 30468 \tnote{d}
\end{tabular} &
\begin{tabular}[c]{@{}l@{}}
1.99999 98196 60040
\\
{}
\\
{}
\\
1.99999 97775 38619
\end{tabular}
\\ \cline{4-5}
$P^{13+}$
& 
\begin{tabular}[c]{@{}l@{}}
1.0150782
\\
0.9992286
\end{tabular}
&
\begin{tabular}[c]{@{}l@{}}
25.1077050
\\
14.4421231
\end{tabular}
&
\begin{tabular}[c]{@{}l@{}}
\textbf{-215.73607} 46141 55184 \tnote{a}
\\
\textbf{-215.73607 56} \tnote{c}
\end{tabular} &
2.00000 13307 34779

\end{tabular}
\begin{tablenotes}
\item[a] {Results obtained using BH$-$ETOs with variational parameters adopted from Ref. \cite{22_Guseinov_2008}}
\item[b] {Results obtained from Ref. \cite{22_Guseinov_2008} }
\item[c] {Numerical HF results obtained from Ref. \cite{23_Koga_1995} }
\item[d] {Results are in agreement with Ref. \cite{22_Guseinov_2008} or improved by re$-$optimization of nonlinear parameters}
\item[*] {Re$-$optimized nonlinear parameters are presented in cases where ground state energy improves; otherwise,  those from Ref. \cite{22_Guseinov_2008} are retained}
\end{tablenotes}
\end{ruledtabular}
\end{threeparttable}
\end{table*}

\section*{Results and Discussions}
The primary objective of this study is to establish the applicability of BH$-$ETOs (Eq. \ref{eq:11}) via Eq. (\ref{eq:18}) for many$-$electron systems. Consequently, calculations are performed for two$-$electron He$-$like ions. Single$-$ and double$-$zeta basis sets are used.\\
Note that test calculations conducted for one$-$electron atoms demonstrate that the optimized quantum numbers take on strictly integer values. Consequently, the standard definition of the Bohr radius remains valid in the non$-$relativistic case, and BH$-$ETOs reduce to Coulomb$-$Sturmian functions. In the case of two$-$electron atoms, BH$-$ETOs, reduce to the standard definition of NSTOs. Given that He$-$like ions possess the $1s^2$ electron configuration, their wavefunctions are characterized by $s-$type orbitals. This conclusion provides direct evidence that the previously hypothesized NSTOs originate from BH$-$ETOs.\\
The radial components of the NSTOs naturally emerge as basis functions in the variational solution of the relativistic Dirac equation. The solution of the Dirac equation for hydrogen$-$like atoms yields the quantum number $\gamma$, which takes non$-$integer values. However, a consistent theoretical framework reconciling the non$-$relativistic and relativistic solutions of hydrogen$-$like atoms has not yet been established. No Dirac$-$like equation solution to date, has been formulated such that highest power of $r$ of its non$-$relativistic limit corresponds to NSTOs. In the non$-$relativistic limit, the Dirac equation solution reduces to the Coulomb$-$Sturmians, wherein the highest power of $r$ corresponds to Slater$-$type orbitals characterized by integer quantum numbers. This issue has recently been addressed in \cite{10_Bagci_2025}, where a Dirac$-$like equation that reduces to BH$-$ETOs in the non$-$relativistic limit has been proposed.

Returning to the central focus of this work, the results for He$-$like ions are presented in Tables \ref{tab:table1}, \ref{tab:table2} and \ref{tab:table3}, where the ground$-$state energies are given. In Tables \ref{tab:table1}, \ref{tab:table2}, and \ref{tab:table3}, the numbers in bold represent values matching the benchmark data. In Tables \ref{tab:table1} and \ref{tab:table2}, the underlined numbers indicate values demonstrating convergence towards the benchmark data. In Table \ref{tab:table3}, the underlined numbers denote values that exactly correspond to those in Ref. \cite{22_Guseinov_2008}. The presented results are consistent with the values obtained in \cite{21_Guseinov_2008, 22_Guseinov_2008, 23_Koga_1995, 24_King_2018, 25_Hatano_2020}, algebraic solution of Hartree$-$Fock equations using NSTOs and Lambda functions, numerical method for solution of Hartree$-$Fock equations respectively, demonstrating the validity of the formalism developed for BH$-$ETOs. As the upper limit of summation increases, the optimized principal quantum numbers obtained in the single$-$zeta approximation, shown in Tables \ref{tab:table1} and \ref{tab:table2}, converge toward integer values $\left( \nu \rightarrow 1 \right)$, thereby confirming Eq. (\ref{eq:17}). The following basis set expansion method is used in the Table \ref{tab:table3},
\begin{align}\label{eq:23}
\psi_{n_{j}l_{j}m_{j}}\left(\vec{r}\right)=
\sum_{i=1}^{2} c_{ij}\chi_{n_{i}^{\ast} l_{i}^{} m_{i}^{} }\left(\zeta^{}_{i}, \vec{r}\right).
\end{align}
This follows from the fact that the electron configuration of He$-$like atoms  $\left(1s^2\right)$ corresponds exactly to a spacial Slater$-$type orbital. The presence of Laguerre polynomials in BH$-$ETOs suggests that calculations for atoms with electron configurations beyond $s$-type may differ from those obtained using NSTOs. A more detailed investigation and comprehensive analysis of the BH$-$ETOs expansion for the Coulomb Green's function (as used in electron scattering from atoms) will be addressed in future work.

\section*{Acknowledgement}
The authors declares no conflict of interest.

\end{document}